\documentclass[12pt]{article}
\usepackage{epsfig}
\usepackage{amsfonts}
\usepackage{amssymb}
\makeatletter
\@addtoreset{equation}{section}

\makeatother

\begin{document}
\vspace{10mm}
\begin{center}
{\Large \bf A Noncommutative Version}
\end{center}

\begin{center}
{\Large \bf of the Minimal Supersymmetric Standard Model}
\end{center}
\vspace{1cm}

\begin{center}
\normalsize
{\large \bf  Masato Arai
\footnote{masato.arai@helsinki.fi},
Sami Saxell
\footnote{sami.saxell@helsinki.fi}
and
Anca Tureanu
\footnote{anca.tureanu@helsinki.fi}
}

\end{center}
\vskip 1.2em
\begin{center}
{\it
High Energy Physics Division,
             Department of Physical Sciences,
             University of Helsinki \\
 and Helsinki Institute of Physics,
 P.O.Box 64, FIN-00014, Finland \\
}
\end{center}
\vskip 1.0cm
\begin{center}
%%%%%%%%%%%%%%%%%%%%%%%%%%%%%%%%%%%%%%%%%%%%%%%%
%
% Abstract
%
%%%%%%%%%%%%%%%%%%%%%%%%%%%%%%%%%%%%%%%%%%%%%%%%

{\large Abstract}
\vskip 0.7cm
\begin{minipage}[t]{14cm}
\baselineskip=19pt
\hskip4mm
A minimal supersymmetric standard model on noncommutative
 space-time (NC MSSM) is proposed.
The model fulfils the requirements of noncommutative gauge
 invariance and absence of anomaly.
The existence of supersymmetry with a scale of its breaking lower than
 the noncommutative scale is crucial in order to achieve a consistent gauge symmetry breaking.
%%%%%%%%%%%%%%%
\end{minipage}
\end{center}

%%%%%%  user's commands  %%%%%%%%%%%%%%%%%%%%%%%%%%%%%%%%%%%%%%%%%%%
\newcommand {\non}{\nonumber\\}
\newcommand {\eq}[1]{\label {eq.#1}}
\newcommand {\defeq}{\stackrel{\rm def}{=}}
\newcommand {\gto}{\stackrel{g}{\to}}
\newcommand {\hto}{\stackrel{h}{\to}}
\newcommand {\1}[1]{\frac{1}{#1}}
\newcommand {\2}[1]{\frac{i}{#1}}
\newcommand{\be}{\begin{eqnarray}}
\newcommand{\ee}{\end{eqnarray}}
\newcommand {\thb}{\bar{\theta}}
\newcommand {\ps}{\psi}
\newcommand {\psb}{\bar{\psi}}
\newcommand {\ph}{\varphi}
\newcommand {\phs}[1]{\varphi^{*#1}}
\newcommand {\sig}{\sigma}
\newcommand {\sigb}{\bar{\sigma}}
\newcommand {\Ph}{\Phi}
\newcommand {\Phd}{\Phi^{\dagger}}
\newcommand {\Sig}{\Sigma}
\newcommand {\Phm}{{\mit\Phi}}
\newcommand {\eps}{\varepsilon}
\newcommand {\del}{\partial}
\newcommand {\dagg}{^{\dagger}}
\newcommand {\pri}{^{\prime}}
\newcommand {\prip}{^{\prime\prime}}
\newcommand {\pripp}{^{\prime\prime\prime}}
\newcommand {\prippp}{^{\prime\prime\prime\prime}}
\newcommand {\pripppp}{^{\prime\prime\prime\prime\prime}}
\newcommand {\delb}{\bar{\partial}}
\newcommand {\zb}{\bar{z}}
\newcommand {\mub}{\bar{\mu}}
\newcommand {\nub}{\bar{\nu}}
\newcommand {\lam}{\lambda}
\newcommand {\lamb}{\bar{\lambda}}
\newcommand {\kap}{\kappa}
\newcommand {\kapb}{\bar{\kappa}}
\newcommand {\xib}{\bar{\xi}}
\newcommand {\ep}{\epsilon}
\newcommand {\epb}{\bar{\epsilon}}
\newcommand {\Ga}{\Gamma}
\newcommand {\rhob}{\bar{\rho}}
\newcommand {\etab}{\bar{\eta}}
\newcommand {\chib}{\bar{\chi}}
\newcommand {\tht}{\tilde{\th}}
\newcommand {\zbasis}[1]{\del/\del z^{#1}}
\newcommand {\zbbasis}[1]{\del/\del \bar{z}^{#1}}
\newcommand {\vecv}{\vec{v}^{\, \prime}}
\newcommand {\vecvd}{\vec{v}^{\, \prime \dagger}}
\newcommand {\vecvs}{\vec{v}^{\, \prime *}}
\newcommand {\alpht}{\tilde{\alpha}}
\newcommand {\xipd}{\xi^{\prime\dagger}}
\newcommand {\pris}{^{\prime *}}
\newcommand {\prid}{^{\prime \dagger}}
\newcommand {\Jto}{\stackrel{J}{\to}}
\newcommand {\vprid}{v^{\prime 2}}
\newcommand {\vpriq}{v^{\prime 4}}
\newcommand {\vt}{\tilde{v}}
\newcommand {\vecvt}{\vec{\tilde{v}}}
\newcommand {\vecpht}{\vec{\tilde{\phi}}}
\newcommand {\pht}{\tilde{\phi}}
\newcommand {\goto}{\stackrel{g_0}{\to}}
\newcommand {\tr}{{\rm tr}\,}
\newcommand {\GC}{G^{\bf C}}
\newcommand {\HC}{H^{\bf C}}
\newcommand{\vs}[1]{\vspace{#1 mm}}
\newcommand{\hs}[1]{\hspace{#1 mm}}
\newcommand{\al}{\alpha}
\newcommand{\Lam}{\Lambda}

\newcommand{\bra}[1]{\langle #1|}
\newcommand{\ket}[1]{|#1\rangle}
\newcommand{\braket}[2]{\langle #1|#2\rangle}

\newcommand{\ms}{M_{\mbox{\tiny SUSY}}}
\newcommand{\me}{M_{\mbox{\tiny eff}}}
\newcommand{\mnc}{M_{\mbox{\tiny NC}}}
\newpage

\section{Introduction}
Quantum field theories (QFT) on noncommutative (NC) space-time
 have been
 subject to intensive research during recent
 years, especially after it
 was shown \cite{sw} that they can be obtained as low-energy limits of
 open string theory in an
 antisymmetric constant background field.
The NC space-time is defined by the commutation relation
\begin{equation}
\left[ \hat{x}^m, \hat{x}^n \right]=i\Theta^{mn}, \label{NC}
\end{equation}
where $\hat{x}^m$ are space-time coordinate operators
and $\Theta^{mn}$ is a constant antisymmetric matrix.
One way to realize field theory on this space-time is to
 replace the usual product
 between any fields with the Moyal star-product
\begin{equation}
(fg)(\hat{x})\longmapsto(f* g)(x) =  e^{\frac{i}{2}\Theta^{mn}
\partial_{x_m}\partial_{y_n}}f(x)g(y)\mid _{x=y}. \label{star}
\end{equation}
This procedure gives rise to new exotic features such as violation of
 Lorentz-symmetry and
 ultraviolet/infrared (UV/IR) mixing
 (for reviews see \cite{szabo, nekrasov}).

Since NC QFT arises as a low energy effective limit from
 string and D-brane theory, it has the potential to provide an
 attractive and motivated framework for physics beyond the standard model
 (SM).
However, it is known that it is difficult to consider a realistic
 phenomenological model building due to a number of
 constraints imposed by noncommutativity.
The main restrictions are from the mathematical (group theoretical) structure
 of the NC gauge theories \cite{terashima,nogo}.
For instance, the only allowed gauge groups in the NC spacetime are
 $U_*(n)$, the NC generalization of the unitary groups $U(n)$.
In Ref. \cite{CPST}, the restrictions imposed by
 noncommutativity were taken advantage of
 and a NC version of the SM
 based on the gauge group $U_*(3)\times U_*(2)\times U_*(1)$ was constructed.
The model can be considered as
 a minimal NC realization of the SM.
Indeed, it leads to the usual SM observable particle content at low
energies. The corresponding symmetry reduction is achieved by the
introduction of
 a scalar field
which was called Higgsac field in Ref. \cite{CPST}, a term which we
shall use also in this paper. In this model,
 the generator of $U(1)_Y$ - the hypercharge group of symmetry - is constructed from a linear combination of
 the generators of the trace-$U(1)$ (in the following we will write simply tr-$U(1)$)
 subgroups of the factors in the
 gauge group $U_*(3)\times U_*(2)\times U_*(1)$.
This model solved the standing problem of electric charge
 quantization observed in \cite{hayakawa}, in which it was shown
 that the only allowed charges for $U_*(1)$ matter are 0 and $\pm 1$,
 and, as a byproduct, it led to all
 the electric charges of the leptons and quarks as unique solution.

While the classical action of this model has many desirable properties,
 it was later found out that it suffers from some serious problems,
 namely violation of unitarity,
 \cite{unitarity}, existence of chiral anomaly \cite{Anomaly,Martin:2001ye}(see also \cite{khozeSM}) and
 problems related the hypercharge
 $U(1)_Y$ sector \cite{MST, Armoni:2001uw, Alvarez-Gaume:2003mb, khoze2}.
Solutions to unitarity and chiral anomaly were proposed in \cite{CKT}
 but the problems with the
 $U(1)_Y$ gauge field remain unsolved.
Taking the one-loop corrections to the polarization tensor of the
 tr-$U(1)$ gauge field into account, the UV/IR mixing effect causes
 an unacceptable infrared singularity.
Furthermore the tr-$U(1)$ gauge field may become tachyonic and one of
 the massless polarizations gets lost.
The latter implies that we observe vacuum birefringence, i.e.,
 a polarization dependent propagation speed.
This leads one to conclude that the $U_Y(1)$ gauge field in the NC SM
 cannot be treated as a photon.
These problems are disastrous for phenomenology.

In an attempt to cure the latter problem, the model was extended
 in Ref. \cite{khozeSM} to the gauge group
 $U_\star(4)\times U_\star(3)\times U_\star(2)$.
The crucial idea behind this extension is
 to construct the $U_Y(1)$ gauge field from a
 traceless combination of $U_*(n)$ generators and to make the tr-$U(1)$ parts
 suffering from the problems mentioned above decouple at low
 energies,
 leaving only a $SU(n)$ symmetry in the low energy effective action.
Indeed, to achieve this purpose, the model in Ref.\cite{khozeSM}
 used the fact that
 the one-loop coupling constant for tr-$U(1)$
 becomes logarithmically smaller as scale decreases.
This running behavior of tr-$U(1)$ is caused by the UV/IR mixing.
However, the running of the tr-$U(1)$ coupling was shown to be too
 slow \cite{khoze2} for complete decoupling, consequently the extra $U(1)$
 gauge field would cause non-acceptable effects at low energies.
In addition, the statement that the UV/IR mixing affects only the
tr-$U(1)$ part and not the "NC $SU(n)$" part is valid only for the
two-point functions ("gluon" propagator). Indeed, the general
picture arrived at in \cite{Armoni:2001uw} is that the UV/IR mixing
effects are given by correlation functions of open Wilson lines,
which implies that 3- and 4-point functions involving "NC $SU(n)$
gluons" exhibit the phenomenon. The gauge $U_\star(n)$ theory was
shown to be renormalizable up to one-loop level \cite{HKT,bonora},
consequently it is not clear how to use the renormalization group
equation and discuss the properties of the $\beta$-function for a
theory whose renormalizability has not been fully proven. The fact
that the leading log approximation works similarly to the
commutative case may be a good educated guess, but may also prove
wrong, since the UV/IR mixing affects
 different diagrams differently
and the dominant diagrams in the UV are not the same as the ones
which are dominant in the IR (and which do not even appear in the
commutative case). We believe that the issue of the UV/IR mixing has
still to be studied, in close connection with the renormalizability.
Though in this paper we give a special attention to the problems of
the tr-$U(1)$ subgroup of $U_*(n)$, which were shown to appear
already at one-loop level, we consider that a clear-cut conclusion
regarding the UV/IR mixing has not yet been reached.

The situation is changed if the theory has supersymmetry (SUSY).
Thanks to SUSY, dangerous quantum corrections in the polarization
  tensor cancel.
As a result, the supersymmetric theory with tr-$U(1)$ gauge group does
 not have the infrared singularity, tachyonic mass nor %and
the polarization
 problem for tr-$U(1)$ gauge field.
The dangerous contribution appears again if SUSY is broken, which is the case
 in a realistic model at low energies.
A theory with soft SUSY
 breaking terms has been studied and it has been shown that
 the infrared singularity actually does not appear \cite{MST} though
 the other problems are
 still left \cite{Alvarez-Gaume:2003mb, khoze2,Carlson:2002zb}.
However, as we argue in this paper the existence of unbroken
supersymmetry at higher scales may suppress these effects enough to
make noncommutative tr-$U(1)$ field a viable candidate for the
photon. It is also well known that in the noncommutative case a
supersymmetric version of the theory has a chance to have no UV/IR
mixing or be renormalizable - an example is the supersymmetric NC
Wess-Zumino model \cite{Girotti:2000gc}. These considerations serve
as an additional motivation for constructing
 a supersymmetrised version of the NC SM.

In this paper, we propose a NC version of the minimal
 supersymmetric standard model (MSSM).
The NC MSSM we construct is based
 on the NC SM of Ref. \cite{CPST}.
We would like to note that the matter content of our model is not
 the same as the commutative MSSM's one.
Requirements that the theory has SUSY, NC gauge invariance and
 cancellation of anomaly lead us to
the introduction of two new extra Higgs fields and two
 leptonic superfields compared to the commutative MSSM matter content.
In addition, in order to achieve the gauge symmetry reduction, we
 introduce a Higgsac superfield which is a supersymmetric extension of
 the Higgsac field proposed in \cite{CKT}.
In the NC setting,
 these fields are inevitably introduced
 in addition to the commutative MSSM matter content, and
 thus we call our model a NC version of MSSM.
Although the NC space-time with the commutation relation (1.1) violates
 the Lorentz invariance, the field theory on such a space-time possesses
 the so-called twisted-Poincar\'e symmetry \cite{CKNT}.
The generators of the latter symmetry satisfy the same algebra as the usual
 generators of the Poincar\'e symmetry.
Thus the representations are
 identical and the particles in NC field theory are still classified by
 their mass and spin. In the case of supersymmetry on NC space-time, a
 twisted version of the super Poincar\'e symmetry
 exists \cite{KS},
 which also justifies the use of the usual particles and their supersymmetric
 partners.
We also briefly discuss the problem of the hypercharge $U_Y(1)$ gauge field
 in our model.
This gauge field is a linear combination of tr-$U(1)$ gauge fields
 and therefore the problems mentioned above appear after SUSY is broken.
We discuss a possible solution to this problem.

There exists also an alternative approach to building a NC
 version of the SM \cite{wessSM} in which the Seiberg-Witten map is
 used to relate the NC gauge theory to a commutative one.
The mapping is based
 on the expansion of the star-products in the Lagrangian.
This allows one
 to write the Lagrangian of a NC version of the SM
 as the Lagrangian of the commutative SM plus an infinite number of
 $\Theta$-dependent terms.
However, this expansion may miss out
 some important NC effects caused by
 the UV/IR mixing.
An approach based on the Seiberg-Witten map would lead to a model different from the one
that we describe in this paper.

The paper is organized as follows.
In section 2 we construct and discuss the minimal supersymmetric version
 of the SM on the NC space-time.
 In section 3 we discuss quantum properties of
 the hypercharge $U_Y(1)$ part
 and SUSY breaking.
Section 4 is devoted to summary and discussion.

\section{NC MSSM}
In this section, we construct the non-commutative version of MSSM.
First of all, we briefly explain the restrictions on model building
in
 non-commutative gauge theory,
 which come from the noncommutativity that constrains the possible gauge groups
 and representations \cite{terashima,nogo}.
As mentioned in the introduction,
 in non-commutative field theory, the only allowed gauge group in NC
 space is the unitary
 group whose Lie algebra is closed under the Moyal bracket
 $[A,B]_*=A*B-B*A$.
The NC unitary group,
 denoted by $U_*(n)$, is obtained by insertion of the star-product
 between the $U(n)$ matrix valued functions.
Especially, it is not possible to have a direct NC
 generalization of $SU(n)$ gauge groups,
 because in NC space the star-product will destroy the closure
 condition. Other restriction is that the charges of the matter fields
 under $U_*(1)$ are quantized to
 just 0, $\pm$1 \cite{hayakawa}.

In addition to these restrictions,
 there is the no-go theorem \cite{nogo} stating that
 the representations of the $u_*(n)$ algebra
 are restricted to $n\times n$ hermitian matrices.
Hence the gauge fields are in $n\times n$ matrix form, while
 the matter fields can only be in fundamental, adjoint or singlet
 states.
Furthermore, matter fields can
 only transform non-trivially under at most two simple subgroups of any
 gauge group consisting of a product of simple groups.
In other words, the matter fields cannot carry more than two NC gauge
 group charges.

The above restrictions cause problems when attempting to construct
NC MSSM. The first restriction tells us that
 one has to start with the gauge group $U_*(3)\times U_*(2)\times
 U_*(1)$ as a minimal extension of the commutative MSSM gauge group
 $SU(3)\times SU(2)\times U_Y(1)$.
The increase in the gauge group implies that there are two
 new additional neutral weak bosons and their superpartners
 in the theory.
These two new states of supermultiplet must be sufficiently massive
 in order to be consistent with present experimental data.
In addition, spontaneous symmetry breaking must take
 place to have the correct MSSM commutative gauge group at low energies.
The second restriction is problematic for construction of the NC MSSM
 based on the gauge group $U_*(3)\times U_*(2)\times U_*(1)$
 since the quarks should have fractional hypercharges.
We also have to pay attention to the last restriction when we make
 charge assignments.
For instance, if the left-handed quark belongs to a fundamental
 representation of $U_*(3)$ gauge group, it should be charged under only one
 of the other groups, i.e., anti-fundamental representation of $U_*(2)$
 and $0$ charge for $U_*(1)$, or singlet for $U_*(2)$ and $-1$ for $U_*(1)$.

In order to break the gauge group $U_*(3)\times U_*(2)\times U_*(1)$ to
 the SM one, its subgroups $U_3(1)\times U_2(1)\times U_1(1)$
 where $U_n(1)$ is tr-$U(1)$ part of $U_*(n)$,
 must be broken down to hypercharge gauge group $U(1)_Y$.
A breaking mechanism was proposed in the construction of the
 NC SM \cite{CPST} by introducing the scalar field charged under
 trace-$U(1)$ group of $U_*(n)$.
This scalar field was called Higgsac. When the Higgsac develops a
vacuum expectation value, the tr-$U(1)$ part of $U_*(n)$ gauge
symmetry is broken. Eventually at non-commutative parameter
$\Theta\rightarrow 0$ limit, the remaining symmetry is $SU(n)$. If
the Higgsac $\phi$ is charged under the $\Theta$-independent
 $U_n(1)\times U_m(1)$-part of $U_*(n)\times U_*(m)$,
 $U_n(1)\times U_m(1)$ gauge group is broken down to a diagonal group $U(1)$.
In the NC SM, two Higgsac fields are necessary to obtain the SM gauge group.
One Higgsac breaks
 $U_3(1)\times U_2(1)$ to a diagonal subgroup $U(1)^\prime$ and the other
 Higgsac then produces a breaking of $U(1)^\prime \times U_1(1)$ to $U(1)_Y$.
Non-zero vacuum expectation values of the two Higgsacs give masses for
 gauge bosons corresponding to the broken $U(1)$ generators while
 massless $U(1)_Y$ hypercharge gauge boson is realized as a linear
 combination of tr-$U_n(1)$.
However, the breaking by this Higgsac field causes a problem on the
 unitarity violation \cite{unitarity}.
The unitarity violation is related to the fact that the symmetry
 reduction by the Higgsac fields is not a spontaneous one, since it transforms
 only under the $\Theta$-independent $U(1)$-part\footnote{In effect, the Higgsac field transforming under the tr-$U(1)$ part of $U_*(n)$ is not an allowed representation
 in the NC case, according to the no-go theorem (see \cite{CKT}) for details.}.
It is not a representation of the gauge group
 and thus the symmetry reduction through Higgsac fields is not a
 spontaneous symmetry breaking mechanism.

In the following we shall explain how the above restrictions are
overcome
 and give complete spontaneous gauge symmetry breaking mechanism in
 construction of the NC MSSM.
Our method is based on Refs. \cite{CPST} and \cite{CKT},
 but we discuss the construction by introducing the superfield on the
 non-commutative superspace.

\subsection{Superfield formalism}
The superfield in the commutative theory is a function of the superspace
 coordinates\footnote{In this paper we follow the notation of Ref. \cite{Wess:1992cp}.}
\begin{eqnarray}
 z^M=(x^m,\theta^\mu,\bar{\theta}_{\dot{\mu}})\,,\,\,\,\mu,\dot{\mu}=1,2\,,
\end{eqnarray}
where $m=0,1,2,3$ is the space-time index, $\theta^\mu(\bar{\theta}_{\dot{\mu}})$ is
 a Grassmann coordinate, and $\mu(\dot{\mu})$ is Weyl
 spinor index.
This superspace is easily generalized to the NC setting \cite{terashima}.
In the NC setting, these coordinates satisfy the following algebra
\begin{eqnarray}
 & [\hat{x}^m,\hat{x}^n]=i\Theta^{mn}, & \\
 & [\hat{x}^m,\hat{\theta}^\mu]=0,& \nonumber \\
 & \{\hat{\theta}^\mu,\hat{\theta}^\nu\}=\{\hat{\bar{\theta}^{\dot{\mu}}},
 \hat{\bar{\theta}^{\dot{\nu}}}\}=\{\hat{\theta}^\mu,\hat{\bar{\theta}^{\dot{\mu}}}\}=0,
 \nonumber &
\end{eqnarray}
where $\hat{\theta}$ and
 $\hat{\bar{\theta}}$ are Grassmann coordinate operators.
The superfield
 is defined just as in the commutative case, and noncommutativity is
 imposed simply by inserting star-products (\ref{star}) into
 the action instead of usual product as
\begin{eqnarray}
 fg(\hat{x},\hat{\theta},\hat{\bar{\theta}})\longmapsto (f*g)(x,\theta,\bar{\theta})=e^{{i\over
  2}\Theta^{mn}\partial_{x_m}\partial_{y_n}}
 f(x,\theta,\bar{\theta})g(y,\theta,\bar{\theta}){\Bigg |}_{x=y}.
\end{eqnarray}
Also the formulation of gauge theories in the superspace is performed just as in the
 commutative case except for star-products between superfields,
  and the restrictions to gauge groups and representations is the same
 as discussed above.

\begin{table}
\begin{center}
\begin{tabular}{cccc}
\hline
Chiral Superfield & $U_\star(3)$ & $U_\star(2)$ & $U_\star(1)$ \\
\hline
$L_i$ & 1 & 2 & 0\\
$\bar{E}_i$ & 1 & 1 & $-1$\\
$Q_i$ & 3 & $\bar{2}$ & 0\\
$\bar{U}_i$ & $\bar{3}$ & 1 & +1\\
$\bar{D}_i$ & $\bar{3}$ & 1 & 0\\
\hline
$L_i^\prime$   & 1 & 2 & $-1$\\
$L_i^{\prime\prime} $ & 1 & 2 & 0\\
\hline
$H_1$ & 1 & $\bar{2}$ & +1\\
$H_2$ & 1 & 2 & $-1$\\
$H_3$ & 1 & 2 & 0\\
$H_4$ & 1 & $\bar{2}$ & 0\\
\hline\\
\end{tabular}
\caption{Matter content of MSSM. The index $i$ denotes the family.}
\end{center}
\end{table}

\subsection{Superpotential}
First we explain matter content and superpotential. After that, we
shall explain how gauge symmetry
 reduction occurs, correct fractional charges for the quarks are
 induced in our model and the anomalies are canceled by introducing new matter fields.
As mentioned earlier, according to the no-go theorem,
 all fields in a NC gauge theory must transform in
 the fundamental, anti-fundamental, adjoint or bi-fundamental
 representation.
We assign the fields to the representation shown in
 Table 1 and construct a superpotential of the NC MSSM.
In the assignment we are guided by the following requirements:
\begin{itemize}
 \item matter content (especially fermions)
 and charge assignment should produce
 the SM (MSSM) hypercharges at low energies,
 \item the theory should be free of anomalies,
 \item the theory (with  matter content and gauge groups) should be minimal.
\end{itemize}
To satisfy these requirements, one is let to the following
superpotential\footnote{There is also another possible choice for
the
 charge assignments of the quarks, leptons and Higgs fields that leads to
 the Standard Model fermion content under symmetry breaking. However,
 this choice requires four additional leptonic doublets to be included
 in order to cancel anomalies, while the the charge assignments
 in Table 1 require only two additional doublets.}:
\begin{eqnarray}
{\cal W}&=&\lambda_e^{ij}H_1 * L_i * E_j+\lambda_u^{ij}Q_i * H_2 * \bar{U}_j
 +\lambda_d^{ij}Q_i * H_3* \bar{D}_j +\mu_{12} H_1 * H_2+\mu_{34} H_3 * H_4 \nonumber \\
&&+\left(
  \alpha_1^{ijk}Q_i*L_j*\bar{D}_k
 +\alpha_2^iL_i*H_4
 +\alpha_3^{i}L_i^\prime*H_1
 +\alpha_4^iL_i^{''}*H_4
 +\lambda_{L^{\prime\prime}}^{ij} H_1 * L^{\prime\prime}_i * E_j
 \right)\,,\nonumber \\ \label{yukawa}
\end{eqnarray}
with the charge assignments as in Table 1. The first
five superfields in Table 1 correspond to the usual
 quarks and leptons of the SM.
In order to give Yukawa terms to all
 the SM fermions
 we introduce an additional Higgs superfield $H_3$ besides the
 superfields $H_1$ and $H_2$ appearing in the commutative MSSM.
This is because
 the superfield $H_1$ giving a down-type mass after electroweak symmetry
 breaking cannot couple to down-type quark due to the charge assignment
 imposed by noncommutativity.
We also introduce the two leptonic chiral superfields $L^\prime$ and
 $L^{''}$, which are necessary to cancel the anomaly as will be discussed below.
The above charge assignments are also required to achieve
 the correct fractional charges for fermions at low energies as
 will be explained.
We also introduce a fourth Higgs to avoid the Witten anomaly (at least
 it is necessary in the $\Theta\rightarrow 0$ limit).
It gives a new $\mu$-term $\mu_{34}H_3*H_4$.

Note that the first two terms inside the parentheses break lepton
number symmetry. In order to remove the effect of these terms we
introduce $R$-parity and
 $R$-parity conservation.
Under $R$-parity, we define
\begin{eqnarray}
 &&L,\bar{E}, Q, \bar{U}, \bar{D}\rightarrow
  -(L,\bar{E},Q,\bar{U},\bar{D})\,, \nonumber \\
 &&L^\prime, L^{\prime\prime} \rightarrow -(L^\prime, L^{\prime\prime})\,,
  \nonumber \\
 &&H_1, H_2, H_3, H_4 \rightarrow H_1, H_2, H_3, H_4\,, \nonumber \\
 &&\theta\rightarrow -\theta\,.
\end{eqnarray}
Then the first four terms inside the parentheses in Eq. (\ref{yukawa}) are removed,
 leaving only one Yukawa term that includes the new leptonic field
 $L^{\prime\prime}$.
However, as will be seen below, the two leptonic fields $L^\prime$ and
 $L^{\prime\prime}$ can
obtain masses through the condensation of the
 Higgsac superfield and thus this term can be neglected at low energies.
Consequently, all the terms inside the parentheses can be dropped out at
 low energies and the low energy superpotential will include only the superfields
 of the commutative MSSM and two additional Higgses $H_3$ and $H_4$.
They give new Yukawa coupling terms (third and fifth terms in Eq. (\ref{yukawa})).
The matter content differs from the commutative
 MSSM by two additional Higgs fields and two leptonic fields.

\subsection{Symmetry reduction}
In this subsection we explain how
 the NC gauge symmetry $U_3(1)\times U_2(1)\times U_1(1)$ which is a
 subgroup $U_*(3)\times U_*(2)\times U_*(1)$ is broken down
 to the hypercharge $U(1)_Y$ gauge group.
As explained earlier, if this breaking is performed by introducing
the
 Higgsac field as in the NC SM \cite{CPST}, which is only charged under tr-$U_n(1)$ subgroup of $U_*(n)$, the
 condensation of the Higgsac field does not mean spontaneous symmetry
 breaking of $U_*(n)$ and problems with unitarity consequently
 arise.
Here we propose a mechanism of truly spontaneous symmetry breaking.

There are two aspects to be considered in a consistent spontaneous
symmetry breaking of the tr-$U(1)$ parts of $U_*(3)\times
U_*(2)\times U_*(1)$. Let one first mention that one Higgsac field
cannot achieve in one step the breaking of all three tr-$U(1)$
subgroups, simply because it does not have enough degrees of freedom
to provide mass for two gauge fields \cite{kibble}. As a result, two
Higgsac fields are needed. If the first is charged under
$U_2(1)\times U_*(1)$, which are broken to a residual $U(1)'$, then
the second has to be charged under $U(1)'\times U_3(1)$. However, at
scales above the first symmetry breaking, this second Higgsac field
is actually charged under $U_3(1)\times U_2(1)\times U_1(1)$,
because the generator of $U(1)'$ is a linear combination of $U_2(1)$
and $U_*(1)$. Thus, the Higgsac fields cannot be constructed without
circumventing the no-go theorem, first because they have to be
charged under such subgroup, that they cannot be representations of
the whole gauge group $U_*(3)\times U_*(2)\times U_*(1)$, and second
because one of the Higgsac fields has to be charged under three
subgroups.

The way of circumventing the no-go theorem is based on the
noncommutative generalization of the gauge invariant operators
using Wilson lines \cite{Gross:2000ba}, leading to the
possibility of constructing tensorial representations of the
noncommutative gauge groups \cite{Chu:2001if}. For the case of
$U_*(3)\times U_*(2)\times U_*(1)$, this approach was initiated in
\cite{CKT}.

\vskip0.5 cm {\it Circumventing the no-go theorem}

For simplicity, first we shall construct a gauge covariant
 Higgsac superfield which breaks tr-$U(1)$ part of $U_*(n)$ gauge group.
In order to construct it, let us first refer to the commutative case
and introduce a chiral superfield
 which is $n$-index antisymmetric representation under
 $U(n)$,
\begin{eqnarray}
 \phi^{[i_1i_2\dots i_n]}(y,\theta)\,, \label{anti-rep}
\end{eqnarray}
which transforms under $U(n)$ as
\begin{eqnarray}
 \phi^{[i_1i_2\dots i_n]}(y,\theta)\rightarrow  (\phi^{[i_1i_2\dots
 i_n]})^U(y,\theta)=U^{i_1}_{i_1^\prime}U^{i_2}_{i_2^\prime}\cdots U^{i_n}_{i_n^\prime}
 \phi^{[i_1^{\prime} i_2^{\prime}\dots i_n^\prime]}
\end{eqnarray}
where $y^m=x^m+i\theta\sigma^m\bar{\theta}$, and its contraction
with epsilon tensor:
\begin{eqnarray}
 \phi(y,\theta)={1 \over n!}\epsilon_{i_1i_2\dots i_n}\phi^{[i_1i_2\dots
  i_n]}(y,\theta)\,, \label{cont}
\end{eqnarray}
which transforms as
\begin{eqnarray}
 \phi(y,\theta)\rightarrow  (\phi)^U(y,\theta)={1 \over n!}\epsilon_{i_1i_2\dots i_n}U^{i_1}_{i_1^\prime}U^{i_2}_{i_2^\prime}\cdots U^{i_n}_{i_n^\prime}
 \phi^{[i_1^{\prime} i_2^{\prime}\dots i_n^\prime]}= (\det U) \phi=
 e^{\mbox{tr} U}\phi\,.
\end{eqnarray}
i.e. the latter chiral superfield has charge $n$ under tr-$U(1)$
 and in the commutative case can cause the breaking of the $U(n)$ gauge group to $SU(n)$ upon condensation.

However, from the no-go theorem, the straightforward noncommutative
generalization of the $n$-index antisymmetric
 object (\ref{anti-rep}) is not an allowed representation of
 $U_*(n)$, because
 $n$ group elements should act from the left:
\begin{eqnarray}\label{tensor}
 \phi^{[i_1i_2\dots i_n]}\rightarrow
 (\phi^{[i_1i_2\dots i_n]})^U
 \equiv U^{i_1}_{i_1^\prime}*U^{i_2}_{i_2^\prime}*\cdots*U^{i_n}_{i_n^\prime}
 *\phi^{[i_1^{\prime} i_2^{\prime}\dots i_n^\prime]}\,,
\end{eqnarray}
where $U$ is a $U_*(n)$ gauge group element.
One can easily see that it does not satisfy group multiplication law, i.e.
\begin{eqnarray}
 ((\phi^{[i_1i_2\dots i_n]})^U)^V=(\phi^{[i_1i_2\dots i_n]})^{V*U}\,. \label{mul}
\end{eqnarray}
Here $V$ is also a gauge group element.
Therefore we cannot treat (\ref{cont}) itself as a representation of $U_*(n)$.

In order to overcome this restriction,
 the proposal of Ref. \cite{Chu:2001if} is to modify the gauge transformation (\ref{tensor}) in a non-trivial, gauge-field-dependent way, so that the group multiplication law holds.
Furthermore one also has to modify (\ref{cont}) to be a $U_*(n)$
gauge
 group representation.
Such a gauge transformation can be constructed
 if the gauge transformation involves the non-commutative version of
 supersymmetric half-infinite Wilson line, $W$.
In the commutative case, the supersymmetric Wilson line is
 constructed in Ref. \cite{SUSYWL}, and then it has been generalized
 to non-commutative setting in Ref. \cite{NCSUSYWL}.
The explicit form for the case of $U_*(n)$ gauge group is given by
\begin{eqnarray}
W&=&P_* \exp_*\left( g \int_0^1 d\sigma {dz^A(\sigma) \over d\sigma}
        A_A\right)\nonumber \\
&=&{\bf 1}_{N}+\sum_{n=1}^\infty {g^n \over n!}\int_0^1 d\sigma_1
 \int_{\sigma_1}^1d\sigma_2\cdots \int_0^1d\sigma_n
 {\partial z^{A_1}(\sigma_1) \over \partial \sigma_1}A_{A_1}*
 \cdots* {\partial z^{A_n}(\sigma_n) \over \partial \sigma_n}
 A_{A_n}. \nonumber \\ \label{NS3-uu}
\end{eqnarray}
Here $A_A$ is the super gauge connection and $z^A=e^A_Mz^M$ are the flat
 superspace coordinates ($A$ runs over Lorentz indices $a$, spinor indices
 $\alpha$ and $\dot{\alpha}$), where $e^A_M$ is a supervielbein.
The supervielbein is
\begin{eqnarray}
e_{M}^{A}\equiv\left(
  \begin{array}{ccc}
\delta^{a}_{m} & 0 & 0 \\
-i\sigma_{\mu\dot{\nu}}^{a}\bar{\theta}^{\dot{\nu}} &
 \delta^{\alpha}_{\mu} & 0\\
-i\theta^\rho\sigma_{\rho\dot{\nu}}^a\epsilon^{\dot{\nu}\dot{\mu}} & 0 &
 \delta^{\dot{\mu}}_{\dot{\alpha}} \label{vielbein} \\
\end{array}
\right)\,,
\end{eqnarray}
and the super gauge connections are given in terms of the unconstrained superfields
 $U$ and $V$ by
\begin{eqnarray}
&A_{\alpha}= -e^{V}D_{\alpha}e^{-V},\hspace{1cm}
 A_{\dot{\alpha}}= -e^{U}\bar{D}_{\dot{\alpha}}e^{-U},\nonumber &\\
&A_{a}=\frac{1}{4}i\bar{\sigma}_a^{\dot{\beta}\alpha}
 \left(-D_\alpha A_{\dot{\beta}}-\bar{D}_{\dot{\beta}} A_\alpha +
 \{A_\alpha,A_{\dot{\beta}}\}\right)\,,&
\end{eqnarray}
where $D_a, D_\alpha$ and $\bar{D}_{\dot{\alpha}}$ are covariant derivatives
 defined by
\begin{eqnarray}
 &\displaystyle e^M_A{\partial \over \partial z^M}\equiv
  D_A=(\partial_a,D_\alpha,\bar{D}_{\dot{\alpha}}), & \nonumber \\
 &\displaystyle D_\alpha={\partial \over \partial
  \theta^\alpha}+i\sigma^m_{\alpha\dot{\alpha}}\bar{\theta}^{\dot{\alpha}}{\partial
  \over \partial z^m},~~~
 \bar{D}^{\dot{\alpha}}={\partial \over \partial
  \bar{\theta}_{\dot{\alpha}}}+i\theta^\alpha\sigma^m_{\alpha\dot{\beta}}
  \epsilon^{\dot{\beta}\dot{\alpha}}{\partial \over \partial z^m}\,.&
\end{eqnarray}
Fixing the gauge partially by demanding $U=0$, the super gauge
 connection can be written purely in terms of the vector superfield $V$:
\begin{eqnarray}
&A_{\alpha}= -e^{V}D_{\alpha}e^{-V},\hspace{1cm}A_{\dot{\alpha}}=0,\nonumber &\\
 & A_{a}=-\frac{1}{4}i\bar{\sigma}_a^{\dot{\beta}\alpha}
 \bar{D}_{\dot{\beta}} A_\alpha.&
 \label{connection}
\end{eqnarray}
Partial gauge fixing leaves the gauge freedom for the vector superfield as
\begin{equation}
e_*^V\mapsto e_*^{-i\Lambda^\dagger}*e_*^V*e_*^{i\Lambda}\,,~~~~
 \bar{D}_{\dot{\alpha}}\Lambda=0\,,
\end{equation}
and thus it can be identified as the superfield whose vector field component is
 the usual NC gauge boson.
The Wilson line operator transforms under gauge transformations as
\begin{equation}
W\mapsto
 e_*^{ig\Lambda(z_1)}*W*e_*^{-ig\Lambda(z_2)},
\end{equation}
where $z_1$ and $z_2$
 are the endpoints of the contour.
As in nonsupersymmetric case, the actual shape of
 the Wilson line is not important.
Then if we choose a half infinite line that starts from infinity
 $z_1=(\infty,\theta,\bar{\theta})$ and
 ends in $z_2=z$ with $\Lambda(z_1)\rightarrow 0$
 this transformation reduces to
\begin{equation}
W(z)\mapsto W(z)*e_*^{-ig\Lambda(z)}\,,
\end{equation}
i.e. the half-infinite Wilson line is an anti-fundamental object
under $U_*(n)$.

Now, by using the supersymmetric half-infinite Wilson line,
 we can modify (\ref{cont}) to be a $U_*(n)$ gauge invariant superfield $\Phi(z)$, while still carrying charge $n$ under its tr-$U(1)$ part\cite{CKT}:
\begin{equation}
\Phi(z)=\frac{1}{n!}\epsilon_{i_1 i_2 ... i_n}
 W^{i_1}_{j_1}(z)*W^{i_2}_{j_2}(z)*...*W^{i_n}_{j_n}(z)*\phi^{[j_1j_2..j_n]}(y)\,,
\label{SPhi}
\end{equation}
where the modified gauge transformation of $\phi^{[i_1i_2..i_n]}$ is
given as in \cite{Chu:2001if}:
\begin{eqnarray}
 \phi^{[i_1i_2\dots i_n]}&\rightarrow &
 (U*W^{-1})_{i_n^\prime}^{i_n}*(U*W^{-1})_{i_{n-1}^\prime}^{i_{n-1}}
 *\cdots *(U*W^{-1})_{i_2^\prime}^{i_2}*(U*W^{-1})^{i_1}_{i_1^{\prime\prime}}
 \nonumber \\
 &&*W^{i_1^\prime}_{i_{1}^{\prime\prime}}*W^{i_2^\prime}_{i_{2}^{\prime\prime}}
 *W^{i_3^\prime}_{i_{3}^{\prime\prime}}*\cdots *W^{i_n^\prime}_{i_n^{\prime\prime}}*
 \phi^{[i_{1}^{\prime\prime} i_{2}^{\prime\prime}\dots i_n^{\prime\prime}]}\,.
\end{eqnarray}
The expression given in Eq. (\ref{SPhi}) is a gauge invariant
Higgsac superfield, and the condensation of
 this superfield causes the spontaneous breaking of tr-$U(1)$ subgroup
 of $U_*(n)$.

Note that for single index representation, say, fundamental
 representation, the gauge transformation law reduces to the one of
 normal non-commutative gauge transformation since the Wilson line
 trivially cancels:
\begin{eqnarray}
 \phi^{i_1}\rightarrow (U*W^{-1})^{i_1}_{i_1^\prime}*W^{i_1^\prime}_{i_1^{\prime\prime}}*\phi^{i_1^{\prime\prime}}=U^{i_1}_{i_1^\prime}*\phi^{i_1^\prime}\,.
\end{eqnarray}

By similar considerations, we can introduce a field charged under an
 arbitrary number of gauge groups, e.g., a field which is in "fundamental
 representation" of $U_*(l)$, $U_*(m)$ and $U_*(n)$, based on the
 auxiliary field  $\phi^{nml}$, with the gauge transformation
\begin{eqnarray}\label{arbitrary}
 \phi^{nml}\rightarrow
  (L*W_L^{-1})^l_{l^\prime}*(M*W_M^{-1})^m_{m^\prime}
 *(N*W_N^{-1})^n_{n^{\prime}}*
 (W_N)^{n^{\prime}}_{n^{\prime\prime}}*(W_M)^{m^{\prime}}_{m^{\prime\prime}}
 *(W_L)^{l^{\prime}}_{l^{\prime\prime}}*
 \phi^{n^{\prime\prime}m^{\prime\prime}l^{\prime\prime}}
\end{eqnarray}
where $l, m$ and $n$ denote the indices for gauge groups $U_*(l), U_*(m)$
 and $U_*(n)$, respectively, $L, M$ and $N$ are corresponding gauge group elements,
 and $W_L$, $W_M$ and $W_N$ are the corresponding half-infinite Wilson
 lines. Although the auxiliary field (\ref{arbitrary}) has a cumbersome transformation
 law and is not in any definite representation of the gauge group $U_*(l)\times U_*(m)\times
 U_*(n)$, it turns out that an object constructed similarly to
 (\ref{SPhi}), i.e.
 $$
\Phi^{[ijk]}=(W_N)^{i}_{n}*(W_M)^{j}_{m}*(W_L)^{k}_{l}*\phi^{[nml]}
 $$
 is invariant under $U_*(l)\times U_*(m)\times
 U_*(n)$.
The introduction of $n$-index field representations and of
representations for
 arbitrary numbers of noncommutative gauge groups by the modification of the gauge transformation implies
 evading the no-go theorem stated in Ref. \cite{nogo}.
A more detailed discussion of this interesting issues is in progress
\cite{AST}.

\vskip0.5cm {\it Symmetry reduction mechanism}

Let us turn to explain how gauge symmetry reduction occurs by the
Higgsac
 superfield (\ref{SPhi}).
In order that $\Phi$ defined in (\ref{SPhi}) develops a vacuum
expectation value, we
 introduce the following superpotential:
\begin{equation}
{\cal W}(\Phi)=m^2\Phi-\frac{\lambda}{3}\Phi*\Phi*\Phi \,.
\end{equation}
Assuming that the perturbative vacuum for the gauge field is given by
 the pure gauge configuration, i.e. $\langle V \rangle =0$, we have
\begin{eqnarray}
 \langle\Phi\rangle=\langle\phi\rangle=\frac{m}{\sqrt{\lambda}}\,. \label{vev}
\end{eqnarray}
In the following we shall see that the tr-$U(1)$ gauge field of $U_*(n)$
 gauge theory has a mass at the perturbative vacuum (\ref{vev}).
First we expand the $\Phi$-field in the NC parameter $\Theta$ and the coupling
 constant $g$:
\begin {eqnarray}
\Phi(z)&=&\left(\det W\right)\phi  \nonumber\\
&=&\left(1+g\int_0^1 d\sigma \frac{dz^A}{d\sigma}\tr A_A
  +\frac{g^2}{2}\int^1_0\!\!\!d\sigma^1\!\!\int^1_0 \!\!\!d\sigma^2
 \frac{dz^A(\sigma^1)}{d\sigma^1} \tr A_A(\sigma^1)\frac{dz^B(\sigma^2)}{d\sigma^2}\;
 \tr A_B(\sigma^2)\right) \phi \nonumber\\
 &&+\,.\,.\,.
\label{expand}
\end {eqnarray}
where
$\phi=\frac{1}{n!}\epsilon_{i_1 i_2 ... i_n}\phi^{[i_1i_2...i_n]}$.
Here the superfields $A_A$ and $\phi$ are $\Theta$-independent and
 the ellipsis denotes terms that are at least of first order in $\Theta_{mn}$ or
 third order in $g$.
Using the Wess-Zumino gauge and equations (\ref{vielbein}) and (\ref{connection}),
 and choosing the contour so that the Grassmann coordinates are constant
 with respect to $\sigma$,
 we can simplify the integrand to
\begin{equation}
{dz^A \over d \sigma}\tr A_A
 =\frac{dz^M}{d\sigma}e^A_M \tr A_A=-n\frac{dx^m}{d\sigma}
 \frac{i}{4}\bar{\sigma}^{\dot{\beta}\alpha}_m\bar{D}_{\dot{\beta}}D_\alpha V^0
\end{equation}
where $V^0$ is the tr-$U(1)$ part of the gauge vector superfield.
Inserting this to the expansion (\ref{expand}) and using the commutation relation
\begin{equation}
\left\{
 D_\alpha,\bar{D}_{\dot{\beta}}\right\}=-2i\sigma^m_{\alpha\dot{\beta}}
 \partial_m\,,
\end{equation}
one can obtain the following form for the kinetic term of the
Higgsac superfield
\begin{eqnarray}
\int d^2\theta d^2\bar{\theta}\Phi^\dagger(z)\Phi(z)=
\int d^2\theta d^2\bar{\theta}
 \phi^\dagger\left(1+ngV^0+\frac{1}{2}(ngV^0)^2\right)\phi
 +\mathcal{O}(\Theta^{mn} )+\mathcal{O}(g^3)
~. \label{kinetic}
\end{eqnarray}
The first three terms in the right hand side represent
 the usual kinetic term for $\phi$ and its
 gauge interactions.
The other terms provide the gauge invariant completion.
It is now clear that if the field $\phi$ has a nonzero vacuum expectation value the
 tr-$U(1)$ gauge boson and its superpartner gaugino have masses.
Decomposing the Lagrangian (\ref{kinetic}), bosonic parts are
\begin{eqnarray}
&&\int d^2\theta d^2\bar{\theta}\Phi^\dagger(z)\Phi(z)=F\bar{F}+\phi \Box
  \bar{\phi}+i\partial_m \bar{\psi}\bar{\sigma}^m\psi
  +ng\left({1\over 2}\bar{\psi}\bar{\sigma}^m\psi
           +{i\over 2}\bar{\phi}\partial_m\phi-{i\over
      2}\partial_m\bar{\phi}\phi\right) \nonumber \\
&&\hspace{20mm}-{ing \over \sqrt{2}}(\phi\bar{\lambda}\bar{\psi}-\bar{\phi}\lambda
 \psi)+{1 \over 2}\left(ng D-{1 \over 2}(ng)^2A_mA^m\right)
 +\mathcal{O}(\Theta^{mn} )+\mathcal{O}(g^3)\,. \nonumber \\ \label{mass-t}
\end{eqnarray}
Substituting
 the vacuum expectation value of $\phi$ (\ref{vev}) into the above equation,
we obtain the following mass terms
\begin{eqnarray}
 {\cal L}_{mass}={i mng \over \sqrt{2\lambda}}(\lambda \psi-\bar{\lambda}\bar{\psi})
 -{g^2 n^2 m^2\over 4 \lambda}A_0^{m}A_{0m}\,,
\end{eqnarray}
where $\lambda, \psi$ and $A_0^m$ are a gaugino, a fermion in the superfield
 $\phi$ and a gauge boson, respectively.
We emphasize that the modified version of the Higgsac mechanism
leads to spontaneous symmetry breaking and thus it should not cause
any problems with unitarity. However, proving this statement
explicitly is highly nontrivial since all the terms in the expansion
of $\Phi$ need to be considered.

Above we described the mechanism in the case of a
 single gauge group  $U_*(n)$.
In the NC MSSM we have to apply this mechanism to
 a direct product of $U_*(n)$ factors, $U_*(3)\times U_*(2)\times U_*(1)$.
In order to obtain SM gauge group, we need to break
 the direct product of gauge groups
 $U_3(1)\times U_2(1)\times U_1(1)$ which is subgroup of
 $U_*(3)\times U_*(2)\times U_*(1)$ to the hypercharge $U(1)_Y$ gauge group.
Its breaking can be achieved by two Higgsac superfields as in
 the NC SM case. In principle, the first Higgsac superfield could be
 charged under any of the three factors, but for the economy of the
 model we shall take it to
 break $U_2(1)\times U_*(1)$ down to $U(1)^\prime$, and then the
 second Higgsac will
 break $U(1)^\prime \times U_3(1)$ down to $U(1)_Y$. This choice is
 motivated by the fact that we have introduced new matter fields,
 $L^\prime$ and $L^{\prime\prime}$, in
 order to obtain a vector-like spectrum under $U_*(2)$ and $U_*(1)$,
as required by the anomaly cancelation condition. The Higgsac field
charged under $U_2(1)\times U_*(1)$ will have also the role to give
masses to the newly introduced matter fields.

Thus, the first composite Higgsac superfield will be
 carrying charge $2$ coupled to tr-$U(1)$ of $U_*(2)$ and charge $-1$
 coupled to $U_*(1)$:
\begin{eqnarray}
\Phi (z)_{U_*(2)\times U_*(1)} &=&\frac{1}{2!} \epsilon_{i_1i_2}
 \left( W_{U_*(2)}\right) _{j_{1}}^{i_{1}}*
 \left( W_{U_*(2)}\right)_{j_{2}}^{i_{2}}*\phi (z)^{[j_{1}j_{2}]}_k
\,,  \label{Higgsac1}
\end{eqnarray}
where $\phi (z)^{[j_{1}j_{2}]}_k$ transforms as
\begin{eqnarray}
 \phi^{[j_1j_2]}_k\rightarrow
 (U_2*W_{U_*(2)}^{-1})_{j_2^\prime}^{j_2}*(U_2*W_{U_*(2)}^{-1})_{j_1^\prime}^{j_1}
*(W_{U_*(2)})^{j_1^\prime}_{j_1^{\prime\prime}}*(W_{U_*(2)})^{j_2^\prime}_{j_{2}^{\prime\prime}}
 *\phi^{j_{1}^{\prime\prime} j_{2}^{\prime\prime}}_k*(U_1^{-1})^k\,,
\end{eqnarray}
where $U_2$ is the element of $U_*(2)$ and $U_1$ is the element of
$U_*(1)$.

The second Higgsac field, charged under tr-$U(1)$ of $U_*(2)$ and
$U_*(3)$ and also under $U_*(1)$ (and after the first symmetry
reduction, under $U(1)^\prime \times U_3(1)$) is given by
\begin{eqnarray}
\Phi (z)_{U_*(3)\times U_*(2)\times U_*(1)} &=&\frac{1}{2!3!}
\epsilon_{i_1i_2}\epsilon^{l_1l_2l_3}
 \left( W_{U_*(2)}\right) _{j_{1}}^{i_{1}}*
 \left( W_{U_*(2)}\right)_{j_{2}}^{i_{2}}*\left( W_{U_*(1)}\right)_{k}\cr
 &*&\phi (z)^{[j_{1}j_{2}k]}_{[n_1n_2n_3]}*\left( W^{-1}_{U_*(3)}\right)^{n_{1}}_{l_{1}}*
 \left( W^{-1}_{U_*(3)}\right)^{n_{2}}_{l_{2}}* \left( W^{-1}_{U_*(3)}\right)^{n_{3}}_{l_{3}}
\,,  \label{Higgsac2}
\end{eqnarray}
where
\begin{eqnarray}
\phi (z)^{[j_{1}j_{2}k]}_{[n_1n_2n_3]}&\rightarrow&
 (U_1*W_{U_*(1)}^{-1})_k*(U_2*W_{U_*(2)}^{-1})_{j_2^\prime}^{j_2}*(U_2*W_{U_*(2)}^{-1})_{j_1^\prime}^{j_1}\nonumber \\
&&*(W_{U_*(2)})^{j_1^\prime}_{j_1^{\prime\prime}}*(W_{U_*(2)})^{j_2^\prime}_{j_{2}^{\prime\prime}}*(W_{U_*(1)})^k*\phi
 (z)^{[j_{1}^{\prime\prime}j_{2}^{\prime\prime}k]}_{[n_1^{\prime\prime}n_2^{\prime\prime}n_3^{\prime\prime}]}\nonumber \\
&& *( W^{-1}_{U_*(3)})^{n_{1}^{\prime\prime}}_{n_{1}^\prime}*
( W^{-1}_{U_*(3)})^{n_{2}^{\prime\prime}}_{n_{2}^\prime}* (
W^{-1}_{U_*(3)})^{n_{3}^{\prime\prime}}_{n_{3}^\prime}\cr &&
*(W_{U_*(3)}*U_3^{-1})_{n_{3}}^{n_{3}^\prime}
*(W_{U_*(3)}*U_3^{-1})_{n_{2}}^{n_{2}^\prime}
*(W_{U_*(3)}*U_3^{-1})_{n_{1}}^{n_{1}^\prime}\,,
\end{eqnarray}
where $U_3$ is an element of the gauge group $U_*(3)$. (The index
$k$ is unnecessary, however we chose to use it in order to show that
the Higgsac fields carry also $U_*(1)$ charge.) Upon the
condensation of these superfields the only tr-$U(1)$ field remaining
 massless is the usual weak hypercharge superfield.

The Higgsac superfield is also used to give masses
 to the doublet leptonic
 superfields $L_i^{\prime}$ and $L_i^{\prime\prime}$.
Indeed, it constitutes the following additional Yukawa couplings in the
 superpotential with the Wilson lines and the composite Higgsac:
\begin{equation}
\left( W_{U_*(2)}*L^{\prime }*W_{U_*(1)}^{-1}\right) ^{T}*\left(
W_{U_*(2)}*L^{\prime \prime }\right) *\Phi _{U_*(2)\times U_*(1)}\,. \label{11}
\end{equation}
The condensation of $\Phi_{U_*(2)\times U_*(1)}$ leads to the
spontaneous
 breaking
 where the surviving $U(1)^\prime$ is a linear combination of tr-$U(1)$
 of $U_*(2)$ and $U_*(1)$.
At the same time the leptonic superfields get a mass of order
 $\langle \Phi_{U_*(2)\times U_*(1)} \rangle$.
Thus, these two superfields do not appear at low energies.
Of course, similar terms could be written also for any other matter
 fields in the theory. Note however that the new leptonic
 fields are vector-like under the SM subgroup of the NC SM
 gauge group.
Thus, decoupling only these fields using the Higgsac field
 does not lead to anomalies for the standard model gauge group.

\subsection{Charge quantization}
The symmetry reduction mechanism proposed here matches exactly, in
the $\Theta\to 0$ limit, the original Higgsac mechanism proposed in
\cite{CPST}. The gauge-covariant completion given by the Wilson
lines insures  just that the symmetry breaking is spontaneous and
problems with unitarity do not appear \cite{AST}. One can check the
fractional charge for quarks in the
 way discussed in \cite{CPST}.
The charge assignments for quarks and the standard model leptons
 in Table 1
 coincide with the ones in the NC SM \cite{CKT},
 so they obviously lead to the correct fractional charges under the
 commutative SM gauge group.

\subsection{Anomaly cancelation}
A solution to the anomaly problem was also given
 in Ref. \cite{CKT}.
In the NC $U_*(n)$ gauge theory, to cancel the anomaly, the
 following conditions should hold \cite{Anomaly}:
\begin{eqnarray}
 &&{\rm Tr}~T^a\{T^b, T^c\}=0\,,\\
 &&{\rm Tr}~T^a[T^b, T^c]=0\,,
\end{eqnarray}
where $T^a$ is a generator of $U_*(n)$ gauge group. The first
equation is the condition for anomaly cancelation for the
 commutative case while the latter is a new condition appearing
 in the NC case.
In order to satisfy these conditions in the NC SM case, it is sufficient
 to introduce two leptonic fields $L^\prime$ and $L^{''}$ whose charge
 assignments are give in Table 1.
For instance, consider the anomaly containing three $U_*(2)$ gauge bosons.
In this case, above conditions are written by
\begin{eqnarray}
&&\sum_f{\rm Tr}~T^a\{T^b, T^c\}=d{\rm Tr}~T^a \{T^b, T^c\}=0\,,\\
&&\sum_f{\rm Tr}~T^a[T^b, T^c]=d{\rm Tr}~T^a [T^b, T^c]=0\,,
\end{eqnarray}
where the sum runs over $U_*(2)$ charged matter and it amounts to the
 constant $d$. Inserting the charge
 assignments given in Table 1
for $L, Q_L, L'$ and $L{''}$, we find for each generation
\begin{eqnarray}
 d=1+(-1)\times 3+1+1=0\,.
\end{eqnarray}
Similarly, one can check the conditions for anomaly cancelation
containing other gauge
 bosons.
Note that we do not have to check the mixed anomaly such as that
 containing two $U_*(3)$ gauge bosons and one $U_*(2)$ gauge boson since
 it does not exist in the NC gauge theory \cite{Martin:2001ye}.
Thus, one concludes that the matter content of the NC SM
 differs from ordinary SM by two additional leptonic
 fields, two Higgsac fields and two additional massive gauge bosons.

The UV/IR mixing which causes problems
 for the hypercharge $U_Y(1)$ gauge field at one-loop level
 will be considered in detail in Section 3 and a possible solution
 in a supersymmetric version of the NC SM will be proposed.

Finally we summarize our ($R$-parity conserving) superpotential:
\begin{eqnarray}
 &&{\cal W}={\cal W}_{\mbox{\tiny Yukawa}}+{\cal W}_{\mbox{\tiny Higgsac}}\,, \\
 &&{\cal W}_{\mbox{\tiny Yukawa}}=\lambda_e^{ij}H_1 * L_i * E_j
   +\lambda_u^{ij}Q_i * H_2 * \bar{U}_j+\lambda_d^{ij}Q_i * H_3* \bar{D}_j
\nonumber \\
&&~~~~~~~~~~~~+\mu_{12} H_1 * H_2+\mu_{34} H_3 * H_4
+\lambda_{L^{\prime\prime}}^{ij} H_1 * L^{\prime\prime}_i *E_j\,,\\
&&{\cal W}_{\mbox{\tiny Higgsac}}=\sum_{a}\left(m^2\Phi_a
 -\frac{\lambda}{3}\Phi_a*\Phi_a*\Phi_a\right)\,,
\end{eqnarray}
where the index $a$ denotes the type of the Higgsac superfield defined
 in (\ref{Higgsac1}) and (\ref{Higgsac2}):
 $a=U_*(2)\times U_*(1), U_*(3)\times U_*(2)\times U_*(1)$ .

\section{Quantum corrections and SUSY breaking}
As we have seen, the model we proposed is supersymmetric, anomaly-free
 and produces the
 correct quantized hypercharges for fermions after NC
 gauge symmetry breaking by super Higgsac fields.
To complete the description of the NC MSSM, we need to specify the
 SUSY breaking.
However, as we mentioned in the Introduction, once SUSY is broken,
 serious problems with the hypercharge $U_Y(1)$ gauge field arise, which
 are caused by the UV/IR mixing.
Let us first
 clarify these problems in detail and then
  discuss a possible solution.

In order to clarify the problems,
 in the following we focus on Euclidean
 SUSY $U_*(1)$ gauge theory
 which was studied
 in Refs. \cite{MST, Carlson:2002zb, Alvarez-Gaume:2003mb, khoze3}.
Recall that $U_Y(1)$ is constructed from the linear combination of
 tr-$U(1)$ parts of the NC gauge groups $U_*(n)$.
A $U_*(1)$ gauge theory involves qualitative features similar to the tr-$U(1)$ factors,
 and so involves all essential features of the problems we explain.
In this setting at the one loop level the polarization tensor for $U_*(1)$ gauge field
 generally has the form \cite{MST,Alvarez-Gaume:2003mb,khoze3}:
\begin{eqnarray}
 \Pi_{mn}=\Pi_1(k^2,\tilde{k}^2)(k^2\delta_{mn}
 -k_m k_n)+\Pi_2(k^2,\tilde{k}^2){\tilde{k}_m \tilde{k}_n
  \over \tilde{k}^2}~~~{\rm with}~~~\tilde{k}^{m}=\Theta^{mn}k_n\,,
\end{eqnarray}
where $k^m$ is the external momentum. The $\Pi_1$ part multiplies
the ordinary transverse structure and is
 related to the gauge coupling by
\begin{eqnarray}
 {1 \over g^2(k^2,\tilde{k}^2)}={1 \over g_0^2}+\Pi_1(k^2,\tilde{k}^2)\,.
\end{eqnarray}
The $\Pi_2$ part is a new Lorentz symmetry violating structure, which is
 specific to NC QFT and explicitly depends on $\Theta_{mn}$.

Performing a one loop calculation for the polarization tensor one
 obtains \cite{Alvarez-Gaume:2003mb}
\begin{eqnarray}
 \Pi_{mn}(k)=\Pi_{mn}(k,l=0)-{\rm Re}\{\Pi_{mn}(k,l=\tilde{k})\}
\end{eqnarray}
with
\begin{eqnarray}
\Pi_{mn}(k,l)&=&2\sum_j\alpha_j \int {d^4q \over (2\pi)^4}\left\{
 d(j)\left[{(2p+k)_m(2_p+k)_n \over (p^2+m_j^2)((p+k)^2+m_j^2)}
 -2{\delta_{mn} \over p^2+m_j^2}
 \right]\right. \nonumber \\
&&~~~~~\left. +4C(j){k^2\delta_{mn}-k_mk_n \over (p^2+m^2_j)((p+k)^2+m_j^2)}
\right\}\exp(ip\cdot l)\,, \label{integral}
\end{eqnarray}
where the coefficients $\alpha_j$, $d(j)$ and $C(j)$ are given in
 Table \ref{coefficients}, and $m_j$ are soft SUSY breaking masses.
Here the $\Pi_{mn}(k,l=0)$ contribution is the so-called planar
 part, while $\Pi_{mn}(k,l=\tilde{k})$ is
 the non-planar part.
The exponential factor $\exp(ip\cdot l)$ in the non-planar part
 gives rise to the UV/IR mixing.
This factor follows from the Moyal star products in the Lagrangian.
At large value of the internal momentum, the exponential factor removes
 the divergence of the integral.
However, when the external momentum goes to zero,
 the divergence reappears.
Thus the divergence at the UV scales is interpreted as an IR singularity.

\begin{table}
\begin{center}
\begin{tabular}{|c|c|c|c|c|}
\hline
j & real scalar & Weyl fermion & gauge boson & ghost \\ \hline
$\alpha_j$ & $-{1 \over 2}$ & ${1 \over 2}$ & $-{1 \over 2}$ & $1$ \\ \hline
$C_j$ & $0$ & ${1 \over 2}$ & $2$ & $0$ \\ \hline
$d_j$ & $1$ & $2$ & $4$ & $1$ \\ \hline
\end{tabular}
\caption{Coefficients in the evaluation of the loop integrals.}
\label{coefficients}
\end{center}
\end{table}

In general, both $\Pi_1$ and $\Pi_2$ are affected by the UV/IR mixing.
In the $\Pi_1$ part, the effect of the UV/IR mixing is to alter the behavior
 of running of the $U_*(1)$ coupling constant.
As $k^2\rightarrow 0$, it has the following form
\begin{eqnarray}
 {1 \over g(k^2,\tilde{k}^2)}=\Pi_1(k^2,\tilde{k}^2)\rightarrow -{b_0 \over
  (4\pi^2)}\log k^2\,,
\end{eqnarray}
while for $k^2 \rightarrow \infty$
\begin{eqnarray}
 {1 \over g(k^2,\tilde{k}^2)}=\Pi_1(k^2,\tilde{k}^2)\rightarrow {b_0 \over
  (4\pi^2)}\log k^2\,,
\end{eqnarray}
where $b_0$ is the one-loop beta-function.

The $\Pi_2$ part arises purely from noncommutativity, but is known to vanish if the theory has exact SUSY \cite{MST}.
On the other hand, for nonsupersymmetric gauge theories the $\Pi_2$ part is nonzero and causes problems.
 The most serious one is that it produces an unacceptable infrared singularity
 $\Pi_2\sim 1/\tilde{k}^2$.
In theories with soft SUSY breaking terms, thanks to the equal
number of bosonic and fermionic degrees of freedom, this infrared singularity
 is cancelled \cite{MST}.
However, a subleading finite term still remains :
\begin{eqnarray}
 \Pi_2\sim \Delta M_{SUSY}^2,~~~\Delta M_{SUSY}^2={1\over 2}\sum_s
  M_s^2-\sum_f M_f^2 \label{sub}
\end{eqnarray}
 unless SUSY is exact.
This produces vacuum birefringence, i.e.
 a mass to only one of the polarizations of the $U_*(1)$
 gauge field, leaving the other polarization massless.
Furthermore, a negative $\Pi_2$ would lead to tachyons while a positive
 mass is phenomenologically strongly constrained.
Consequently, it would seem that the $U_Y(1)$ gauge field, which is
 a linear combination of tr-$U_*(1)$ fields, has serious problems already at one-loop level \cite{Armoni:2001uw,khoze2}.

One way to avoid these problems is to modify the
 UV physics \cite{Abel:2006wj}.
In the above analysis, it is assumed that the NC QFT is valid up to
 arbitrarily large momentum scales.
However, if NC QFT is realized as a low energy
 effective theory of some
 underlying theory such as string theory, the theory should be
 modified above some UV scale,
 e.g. string scale.
Indeed, since NC QFT is realized as a special limit of open strings
 in a background of antisymmetric tensor field $B_{mn}$,
 it is expected that at least above the string scale the $B_{mn}$ does not have a
 vacuum expectation value and the noncommutativity does not appear there.
This modification affects the low energy physics since the infrared
 region receives the effect from the UV domain due to the UV/IR mixing.
The modification actually improves the situation,
 leading to the birefringence effect.
In Ref. \cite{Abel:2006wj}, it is shown that constraints on this effect
 require the noncommutativity scale to be close to the Planck scale.
Furthermore, in this setting
 the behavior of the running coupling constant is exactly
 the same with the commutative one at the low energies smaller than some
 infrared scale specified by
 $\Lambda_{\mbox{\tiny \rm IR}}\sim \Lambda_{\mbox{\tiny \rm UV}}^2/\mnc$.
This is a desired property for considering phenomenological model
 building although the photon may be still tachyonic (the tachyonic mode is
 possible since the Lorentz symmetry is now broken).

Here we briefly discuss another possibility for improving the
 above situation.
We assume  that SUSY in the theory is spontaneously broken
 by a mechanism such as
 O'Raifeartaigh and Fayet-Iliopoulos mechanisms.
The breaking of SUSY is assumed to occur at a hidden sector, so that
 its effect on the visible MSSM sector is the occurrence of soft SUSY
 breaking masses $m_i$ below some SUSY breaking scale $\ms$.
Above this scale, SUSY is expected to be restored
 and all the soft SUSY breaking terms to vanish.
Recalling that the $\Pi_2$ part vanishes in theories with exact SUSY,
 it is expected that the above situation renders the $\Pi_2$ contribution
 small so that the one-loop correction to the polarization tensor
 is consistent with the experimental bound for the
 Lorentz violation.
To be more precise
 we also need to consider the relation between $\ms$ and
 the NC scale $\mnc\sim |\Theta|^{-1/2}$.

In order to estimate the one-loop contribution to $\Pi_2$ in the
 setting mentioned above,
 we divide the integral (\ref{integral}) into two parts
\begin{eqnarray}
\Pi_{mn} &=&(\Pi_a)_{mn}+(\Pi_b)_{mn}\nonumber \\
&=&\left(\int_{0 \le |p| \le \ms}(m_j\neq 0)
+\int_{\ms \le |p| \le \infty}(m_j=0)\right){d^4 p \over
 (2\pi)^4}\cdots \,, \label{int1}
\end{eqnarray}
where the ellipsis denotes the integrand.
Here we treat the soft SUSY breaking mass term as a step
 function like, i.e. $m_j=0~(\ms \le |p| \le \infty)$ and
 $m_j\neq 0~(0 \le |p| \le \ms)$.
Note that in this treatment gauge invariance gets lost
 since $\ms$ behaves as a cut-off which leads to a non-gauge invariant
 term in the calculation.
However, we simply ignore the non-gauge invariant term in what follows.
Our purpose is to see whether this setting can solve the
 problem of tr-$U(1)$ part.
We believe that this qualitative argument does not change in
 a proper gauge invariant regularization and
we postpone a more rigorous analysis to a future work.
The first integral $(\Pi_a)_{mn}$ in (\ref{int1}) can be written as
 \cite{Alvarez-Gaume:2003mb}:
\begin{eqnarray}
 (\Pi_a)_{mn}(k)&=&{1 \over 4\pi^2}(k^2\delta_{mn}-k_m k_n) \nonumber \\
 &&\times \sum_j \alpha_j \int_0^1 dx[4C(j)-(1-2x)d(j)]
   \left[K_0\left({\sqrt{A_j} \over \ms}\right)
        -K_0\left({\sqrt{A_j} \over \me}\right)\right] \nonumber \\
 &+&{1 \over 4\pi^2}\tilde{k}_m\tilde{k}_n \me^2
    \sum_j\alpha_jd(j)\int_0^1 dx A_j K_2\left({\sqrt{A_j} \over
                      \me}\right) \nonumber \\
 &+&\delta_{mn}[\mbox{gauge~non-invariant~term}]\,, \label{int2}
\end{eqnarray}
where
\begin{eqnarray}
 A_j=m_j^2+x(1-x)k^2,~~~~~{1 \over \me^2}={1 \over \ms^2}+\tilde{k}^2\,.
\end{eqnarray}
The non-planar part in the second integral $(\Pi_b)_{mn}$ in (\ref{int1})
 is exactly zero
 because SUSY is manifest there and therefore only the first integral
 will contribute to the birefringence effect.
Using (\ref{int2}) one can obtain the following approximate expressions
 for $\Pi_2$ part from (\ref{int1}) \cite{Abel:2006wj}
\begin{eqnarray}
 \Pi_{2}=
\left\{
\begin{array}{l}
 D\Delta \ms^2 \,,~~~~~~\mbox{for}~~\displaystyle {\mnc^2 \over
  \ms}\ll k \ll \Delta \ms\,, \\
 D^\prime \Delta \ms^2\ms^2 \tilde{k}^2\,,~~~~~\mbox{for}~~
 \displaystyle k \ll {\mnc^2 \over
  \ms}\,,~~~m_j^2\ll \ms^2\,, \label{pi2}
\end{array}
\right.
\end{eqnarray}
where $D$ and $D^\prime$ are known constants.
One can see that the second equation in (\ref{pi2}) is suppressed by the
 NC scale $\mnc$ while the first equation does not have any suppression.
In the following we consider only the second
expression, which indeed gives a promising result.

Following the discussion in Ref. \cite{khoze2}, we consider the
 equation of motion for the photon,
\begin{eqnarray}
 \Pi^{mn}(k)A_n(k)=0\,. \label{eq}
\end{eqnarray}
We specify the noncommutativity as
 $\Theta^{13}=-\Theta^{31}=\Theta=1/\mnc^2$, while all other components
 of $\Theta^{mn}$ are taken to be zero.
The photon propagates
in the third direction $k^\mu=(k^0,0,0,k^3)$.
Now we have two polarization vectors
\begin{eqnarray}
 A_1^m=(0,1,0,0),~~~A_2^m=(0,0,1,0)\,.
\end{eqnarray}
Substituting this into Eq. (\ref{eq}), we find
\begin{eqnarray}
 (\Pi_1k^2-\Pi_2)A_1^m&=&0\,, \label{photon1} \\
 \Pi_1 k^2 A_2^m&=&0\,. \label{photon2}
\end{eqnarray}
The second equation representing the equation of motion for a
polarized
 photon along $A_2^m$ behaves like ordinary photon while the first
 equation for polarized photon $A_1^m$ receives a new effect, $\Pi_2$.
Now in order to study in more detail we substitute the expression
 (\ref{pi2}) into (\ref{photon1}).
Then we obtain
\begin{eqnarray}
k^2+D^\prime{\Delta \ms^2 \ms^2 \over \Pi_1 \mnc^4}(k^3)^2=0\,, ~~~~~\mbox{for}~~
 \displaystyle k \ll {\mnc^2 \over
  \ms}\,,~~~m_j^2\ll \ms^2\,. \label{suppress}
\end{eqnarray}

Now we consider the dispersion relation of photon for the case
(\ref{suppress}).
Restoring the light speed $c$ and using $k^0=\omega$ for the frequency of
 the wave, we find
\begin{eqnarray}
 \omega^2-c^2\left({1 \over 1+\Delta n}\right)^2(k^3)^2=0\,,
\end{eqnarray}
with
\begin{eqnarray}
 \Delta n&\simeq& {D^\prime \over 2 \Pi_1}{\Delta \ms^2 \ms^2
  \over \mnc^4}\,,
\end{eqnarray}
where $D^\prime=1/4\pi^2$.
In Ref. \cite{Kostelecky:2002hh}, all possible dimension four Lorentz
 violating operators in electrodynamics were studied and the constraints
 were obtained.
The Lorentz violating operators
 can be related to our $\Delta n$ \cite{Abel:2006wj}.
Most strongest bound in Ref. \cite{Kostelecky:2002hh} is
 obtained from observation of objects at cosmological distances
\begin{eqnarray}
 |\Delta n_{\mbox{\tiny cosmo}}|\le 10^{-37}-10^{-32}\,.
\end{eqnarray}
Our result is consistent with this bound, for instance, if we take the
 following values for scales: $\ms\sim 10^{10}, \mnc\sim 10^{18},
 m_j\sim 10^2$ and $k \sim 100$ GeV.
In the above setting, the $\Delta n$ is found
 to be
\begin{eqnarray}
 \Delta n \sim 10^{-62}\,.
\end{eqnarray}
Our setting is consistent with the experimental bound for the Lorentz
 violation although the photon may be tachyonic similarly to the case in
 Ref. \cite{Abel:2006wj}.

We briefly summarize our argumentation. At scales above $M_{\mbox{\tiny SUSY}}$, SUSY
is exact and thus the contribution to $\Pi_2$ from these scales is
exactly cancelled. Furthermore, if $M_{\mbox{\tiny SUSY}}$ is much smaller than the scale of noncommutativity,
the contribution from scales below $M_{\mbox{\tiny SUSY}}$ is expected to be negligible because in this region the theory
should be effectively commutative.

Note that in our case the behaviour of $U(1)$
 running coupling constant is still altered by the UV/IR mixing
 compared to the commutative one,
 while in the theory with the UV completion in
 \cite{Abel:2006wj} the running behaviour is the same with the
 commutative one below the scale
 $\Lambda_{\mbox{\tiny IR}}\sim \Lambda_{\mbox{\tiny UV}}^2/\mnc$.
We leave a further study on the behaviour of the running coupling constant for later work.

\section{Summary and discussion}
In this paper, we have proposed a NC version of MSSM.
The guiding principles in constructing the action are to require SUSY, NC
 gauge invariance, absence of anomalies and correct
 fractional $U_Y(1)$ charges for fermions.
The NC gauge invariance requires to introduce two extra gauge fields and
 their superpartners
 compared to the MSSM gauge field content, since our model is based
 on the NC gauge groups $U_*(3)\times U_*(2)\times U_*(1)$.
Other requirements lead us to
 introducing
 two new Higgs superfields $H_3, H_4$
 for the construction of the Yukawa couplings and two leptonic superfields
 $L^\prime, L^{''}$ for the cancellation of the
 anomaly, compared to the commutative MSSM matter content.

Further additional matter, the Higgsac superfield, is introduced.
This plays two roles in our model.
One of them is to reduce the NC gauge symmetry from
 $U_*(3)\times U_*(2)\times U_*(1)$ to the SM one.
The two extra gauge bosons and their
 superpartners are decoupled at low energies via the
 condensation of
the Higgsac superfield, which gives masses to the extra gauge fields.
The other role is to achieve the decoupling of the extra leptonic
 superfields.
The Higgsac superfield can form Yukawa coupling with these two leptonic superfields
 through the NC Wilson lines.
Upon the condensation of the Higgsac, the leptons become massive and thus
 they decouple at low energies.

We have discussed the quantum properties of our model, especially concerning
 the hypercharge $U_Y(1)$ sector.
In the NC SM proposed in Ref. \cite{CPST}, the hypercharge $U_Y(1)$
 gauge field suffers from the problems on the tachyonic mass and
 the vacuum birefringence once the one-loop corrections are taken into account.
The $U_Y(1)$ gauge group is a linear combination of three
 tr-$U(1)$ gauge groups which are subgroups of the
 $U_*(n)$ gauge groups.
In the NC setting, the tr-$U(1)$ gauge field is affected by the UV/IR
 mixing and it generates a Lorentz violating term in the polarization
 tensor.
As a result, tr-$U(1)$ gauge field has the serious problems mentioned
 above.
In this paper, we gave a possible solution to this problem, by assuming
 that supersymmetry is restored
 above the SUSY breaking scale $\ms$.
Then, with appropriate values of the scales
 $\ms, \mnc$ and $m_j$, the Lorentz violating term in the polarization tensor
 has small enough value to avoid violating experimental bounds.

Note that our model does not go to the commutative MSSM in the limit of
 vanishing noncommutativity parameter $\Theta\rightarrow 0$ at the tree level since
we introduced two new Higgs superfields $H_3$ and $H_4$ and the interactions
 with these fields do not vanish in this limit (other terms
 including the new leptonic fields do not appear at low energies as mentioned above).
This implies that our model involves many interesting new phenomenological
 features compared to the commutative MSSM.

One of the most important things to study
 in the low energy physics of our model is the
 electroweak symmetry breaking.
In our model, since a new down-type Higgs $H_3$ is introduced beside
 $H_4$, these fields must obtain
vacuum expectation values in order to give masses
 to down-type quarks while $H_1$ and $H_2$ corresponding to the usual Higgses
 appearing in the commutative MSSM %also
 have to condensate in order to give masses
 to up-type quarks and leptons.
In the commutative MSSM with soft SUSY breaking terms, electroweak
 symmetry breaking does not occur at the tree level, but taking one-loop
 corrections to Higgs mass into account, the Higgs mass runs in the
 infrared and eventually goes to negative \cite{Inoue:1982ej}.
As a result, the Higgs field has a nonzero vacuum expectation value and
 electroweak symmetry breaking occurs.
This mechanism may be also applied in our model.
We postpone a more detailed study of this issue for future work.
It would also be interesting to extend NC version of the SM
 \cite{wessSM} using
 Seiberg-Witten map to the supersymmetric case.

\vspace{10mm}
%%% Acknowledgements  %%%%%%
\noindent {\Large \bf Acknowledgements}\\\\
We are grateful to Masud Chaichian for illuminating discussions and
 careful reading of the manuscript.
We thank Adi Armoni, Valentin V. Khoze, Archil Kobakhidze and M. M.
Sheikh-Jabbari for useful correspondence. The work of M.A. is
supported by the bilateral program of Japan Society
 for the Promotion of Science and Academy of Finland, ``Scientist Exchanges''.
S.S. acknowledges a grant from GRASPANP, the Finnish Graduate School
 in Particle and Nuclear Physics.

%%%%%%%%%%%%%%%%%%%%%%%%%%%%%%%%%%%%%%%%%%%%%%%%%

\end{document}